\documentclass[%
 reprint,
superscriptaddress,
 amsmath,amssymb,
 aps,
]{revtex4-2}

\usepackage{graphicx}% Include figure files
\usepackage{dcolumn}% Align table columns on decimal point
\usepackage{bm}% bold math
\usepackage{makecell}
\usepackage{hyperref}
\usepackage{cleveref}
\begin{document}

\preprint{APS/123-QED}

\title{Enhancing Events in Neutrino Telescopes through Deep Learning-Driven Super-Resolution}% Force line breaks with \\

\author{Felix J. Yu}
%  \altaffiliation[Also at ]{Physics Department, XYZ University.}%Lines break automatically or can be forced with \\
\email{felixyu@g.harvard.edu}
\affiliation{The NSF AI Institute for Artificial Intelligence and Fundamental Interactions}
\affiliation{Department of Physics and Laboratory for Particle Physics and Cosmology, Harvard University, Cambridge, MA 02138, US}

\author{Nicholas Kamp}
\email{nkamp@g.harvard.edu}
\affiliation{Department of Physics and Laboratory for Particle Physics and Cosmology, Harvard University, Cambridge, MA 02138, US}

\author{Carlos A. Arg\"{u}elles}
\email{carguelles@g.harvard.edu}
\affiliation{The NSF AI Institute for Artificial Intelligence and Fundamental Interactions}
\affiliation{Department of Physics and Laboratory for Particle Physics and Cosmology, Harvard University, Cambridge, MA 02138, US}

\date{\today}% It is always \today, today,
             %  but any date may be explicitly specified

\begin{abstract}
Recent discoveries by neutrino telescopes, such as the IceCube Neutrino Observatory, relied extensively on machine learning (ML) tools to infer physical quantities from the raw photon hits detected.
Neutrino telescope reconstruction algorithms are limited by the sparse sampling of photons by the optical modules due to the relatively large spacing ($10-100\,{\rm m})$ between them.
In this letter, we propose a novel technique that learns photon transport through the detector medium through the use of deep learning-driven super-resolution of data events.
These ``improved'' events can then be reconstructed using traditional or ML techniques, resulting in improved resolution.
Our strategy arranges additional ``virtual'' optical modules within an existing detector geometry and trains a convolutional neural network to predict the hits on these virtual optical modules.
We show that this technique improves the angular reconstruction of muons in a generic ice-based neutrino telescope.
Our results readily extend to water-based neutrino telescopes and other event morphologies.
\end{abstract}

%\keywords{Suggested keywords}%Use showkeys class option if keyword
                              %display desired
\maketitle

% \tableofcontents

% For Letter: Keep Figure 1, 3, 6, move rest to appendix

\section{\label{sec:intro}Introduction}

Neutrino telescopes have opened a new window into the Universe beyond the energies and distances available to radio and gamma-ray astronomy, respectively~\cite{Arguelles:2024xkx}.
This is facilitated by the largest particle detectors ever built, typically an array of light detectors distributed throughout a gigaton of water or ice.
Neutrino astronomy is currently led by the IceCube Neutrino Observatory, which has instrumented a cubic kilometer of South Pole ice with 5160 digital optical modules (DOMs)~\cite{IceCube:2016zyt} arranged in 86 vertical strings, such that DOMs are spaced by 17\,m (125\,m) in the vertical (horizontal) direction.
IceCube has made a series of breakthrough discoveries over the past decade, including the first observation of high-energy astrophysical neutrinos~\cite{IceCube:2013cdw,IceCube:2013low} as well as the first neutrinos from a blazar, TXS~0506+056~\cite{IceCube:2018cha}, and an active galaxy, NGC~1068~\cite{IceCube:2022der}.
More recently, IceCube made the first detection of neutrinos from our own galactic plane~\cite{IceCube:2023ame}, relying extensively on machine learning (ML) tools to determine the energy and direction of the original neutrino~\cite{Abbasi:2021ryj,IceCube:2021umt}.
This is only one example of the emerging importance of ML techniques in neutrino telescopes~\cite{IceCube:2022njh, Kharuk:2023xnl, Eller:2023myr, DeSio:2019lcr, Mo:2023rgb, Reck:2021zqw, IceCube:2018gms, Yu-2024, Jin:2023xts, Zhu:2024ubz, IceCube:2021umt} and neutrino physics as a whole~\cite{nature-ml, Li:2022frp, Imani:2023blb, MicroBooNE:2018kka, MicroBooNE:2020yze, Aurisano:2016jvx, MicroBooNE:2020yze, Domine:2019zhm, Drielsma:2021jdv}.

Building upon the success of IceCube, several new ice-and water-based neutrino telescopes are being built or planned.
In the former category are KM3NeT in the Mediterranean Sea~\cite{KM3NeT:2018wnd}, Baikal-GVD in Lake Baikal~\cite{Baikal-GVD:2018isr}, and P-ONE in the Pacific Ocean~\cite{P-ONE:2020ljt}.
In the latter category are IceCube-Gen2~\cite{IceCube-Gen2:2020qha} and a series of water-based detectors in China, including TRIDENT~\cite{Ye:2022vbk} and HUNT~\cite{Huang:2023mzt}, all of which would increase the instrumented volume by an order of magnitude.
Due to the larger size, the light sensors in these next-generation detectors will be spaced further apart.
This will reduce the information available for each detected neutrino interaction, especially at lower energies.

\begin{figure*}
    \centering
    \includegraphics[width=0.95\textwidth]{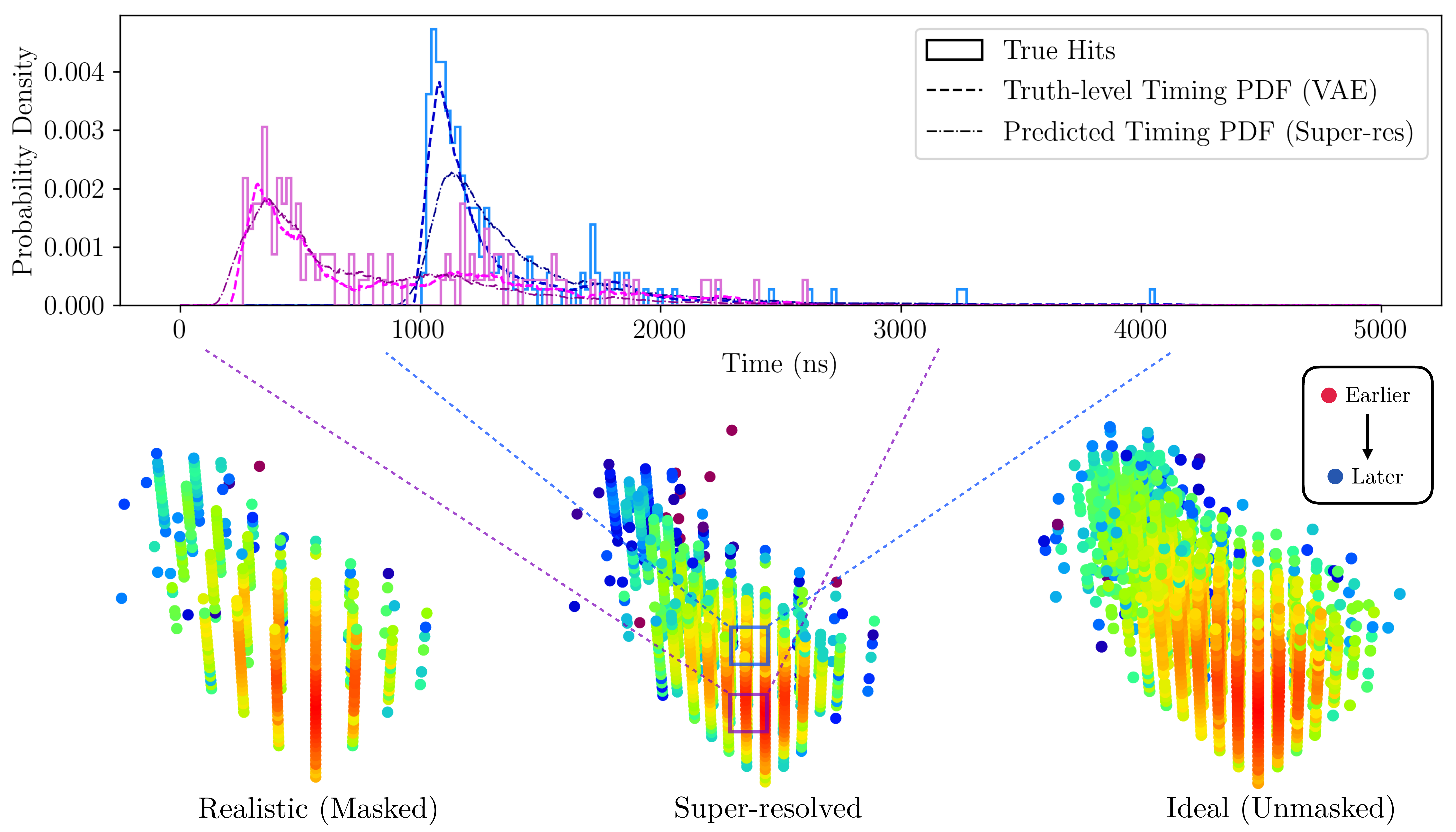}
    \caption{\textbf{\textit{Event displays, showing the masked, super-resolved and unmasked event.}} The unmasked and masked events are obtained from simulation, representing ideal and realistic detector configurations, while the super-resolution network attempts to enhance the masked event into the unmasked. The top plot shows the photon arrival time series from the super-resolution network and the pre-trained VAE on two particular virtual OMs in the super-resolved event.}
    \label{fig:eventdisplay}
\end{figure*}

To address this, we introduce the concept of \textbf{super-resolution} for neutrino telescope data.
This technique is used in the image-processing community to increase the definition of the pictures at the sub-pixel level~\cite{wang2020deep}.
For neutrino telescope data, super-resolution translates to sampling photon depositions between optical modules (OMs).
In this letter, this is achieved by training a deep-learning algorithm to predict detected photons on virtual OMs arranged between the physical OMs, as depicted in \cref{fig:eventdisplay}.
We show that angular reconstruction algorithms perform better on super-resolved events with information from the virtual DOMs.
This will be essential for improving the sensitivity of existing and future neutrino telescopes to astrophysical neutrino sources within and beyond our galaxy.
Further, the techniques proposed in this letter can improve particle reconstruction in other light-based neutrino experiments, such as Super-Kamiokande~\cite{Super-Kamiokande:2002weg}, Hyper-Kamiokande~\cite{Hyper-Kamiokande:2018ofw}, JUNO~\cite{JUNO:2021vlw}, and THEIA~\cite{Theia:2019non}, among others.

The rest of the letter is organized as follows.
\Cref{sec:arch} describes the architecture of the super-resolution network.
Next, \cref{sec:events} discusses the training procedure and dataset generation.
\Cref{sec:results} demonstrates the improved angular resolution for super-resolved muon events in an IceCube-like detector.
Finally, \cref{sec:conclusion} summarizes the results and potential future directions of this work.
The code implemented for this letter is made available at \cite{GithubCode}.

\section{Architecture} \label{sec:arch}

The basic premise of the super-resolution method is to start by simulating events within a denser geometry and then mask a portion of these strings during training.
The network is trained to reconstruct the sensors on the masked strings, which we will refer to as virtual strings. 
Our architecture is independent of the specific detector geometry, while simultaneously preserving as much timing information as possible.
This versatility enables the network to adapt to various geometries and detectors, as well as to different event morphologies.

\begin{figure*}
    \centering
    \includegraphics[width=0.95\textwidth]{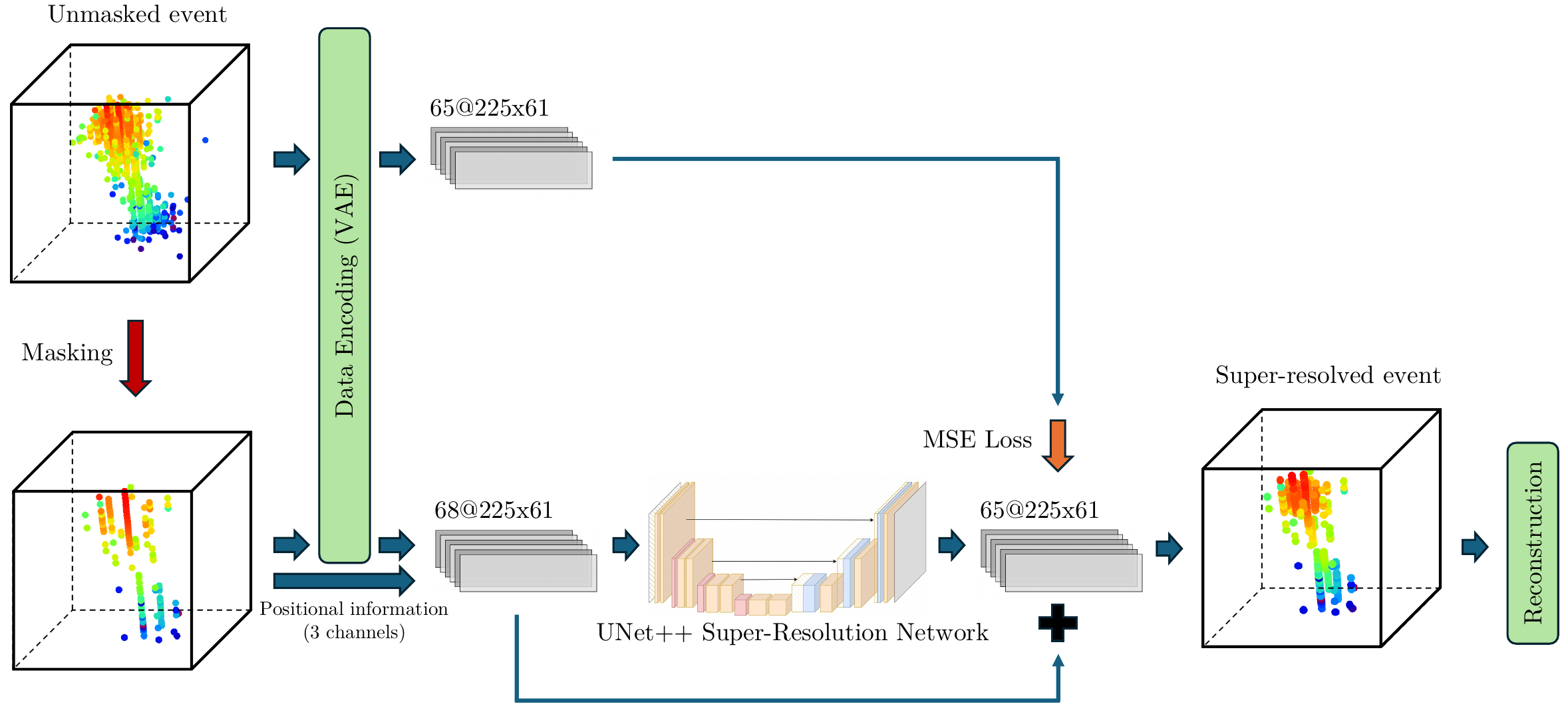}
    \caption{\textbf{\textit{Pipeline of the super-resolution framework.}} The OM timing information is encoded into a 64-parameter latent vector. Neutrino telescope events are arranged into 2D images by string and sensor on a string. For the network inputs, each OM contains 68 features: 3 (3D sensor position) + 1 (number of photon hits) + 64 (timing latent vector).}
    \label{fig:architecture}
\end{figure*}

Our implementation of super-resolution is organized by a high-level algorithm that compares predicted super-resolved events with their true, unmasked events, as depicted in \cref{fig:architecture}.
This is achieved by encoding each event into a 2D image and passing these images through a UNet-type network.
The output is also a 2D image representing the super-resolved event.
One of the main challenges with this pipeline is adequately encoding and decoding the timing information, which is needed for the prediction of the time series on virtual OMs.
This difficulty arises from the geometrically sparse distribution of photon hits for any given neutrino interaction and the highly variable length of the photon hit time series in a given OM.
Recently, Ref.~\cite{IceCube:2021umt} demonstrated that an asymmetric Gaussian mixture model can effectively represent the time series of shower-like events in IceCube.
Building upon this concept, we train a variational autoencoder (VAE) to encode the time series into a compressed latent space~\cite{Yu-2024}.
The core idea of this process is to take a finely binned distribution of photon hit times (we chose to use 5000 bins, each 1 ns wide) and compress it into a more compact yet effective representation—a 64-parameter latent vector.
The UNet then operates fully in the latent space, taking in a point in the latent space for each physical OM and predicting one for each virtual OM.
Examples of the VAE representation and super-resolution network predicted time series for two specific virtual OMs are shown in the top plot of \cref{fig:eventdisplay}.
While the shapes of most predicted timing PDFs closely match that of the true hits, those farther from the interaction point, with fewer statistics, are harder to reconstruct accurately. 
Given the limited data on these OMs, it is often sufficient to capture key features like the first hit time and the position of the peak, rather than the exact shape of the distribution.

\section{\label{sec:events}Event Generation and Training}

\subsection{Virtual Strings}

\begin{figure}
    \centering
    \includegraphics[width=0.48\textwidth]{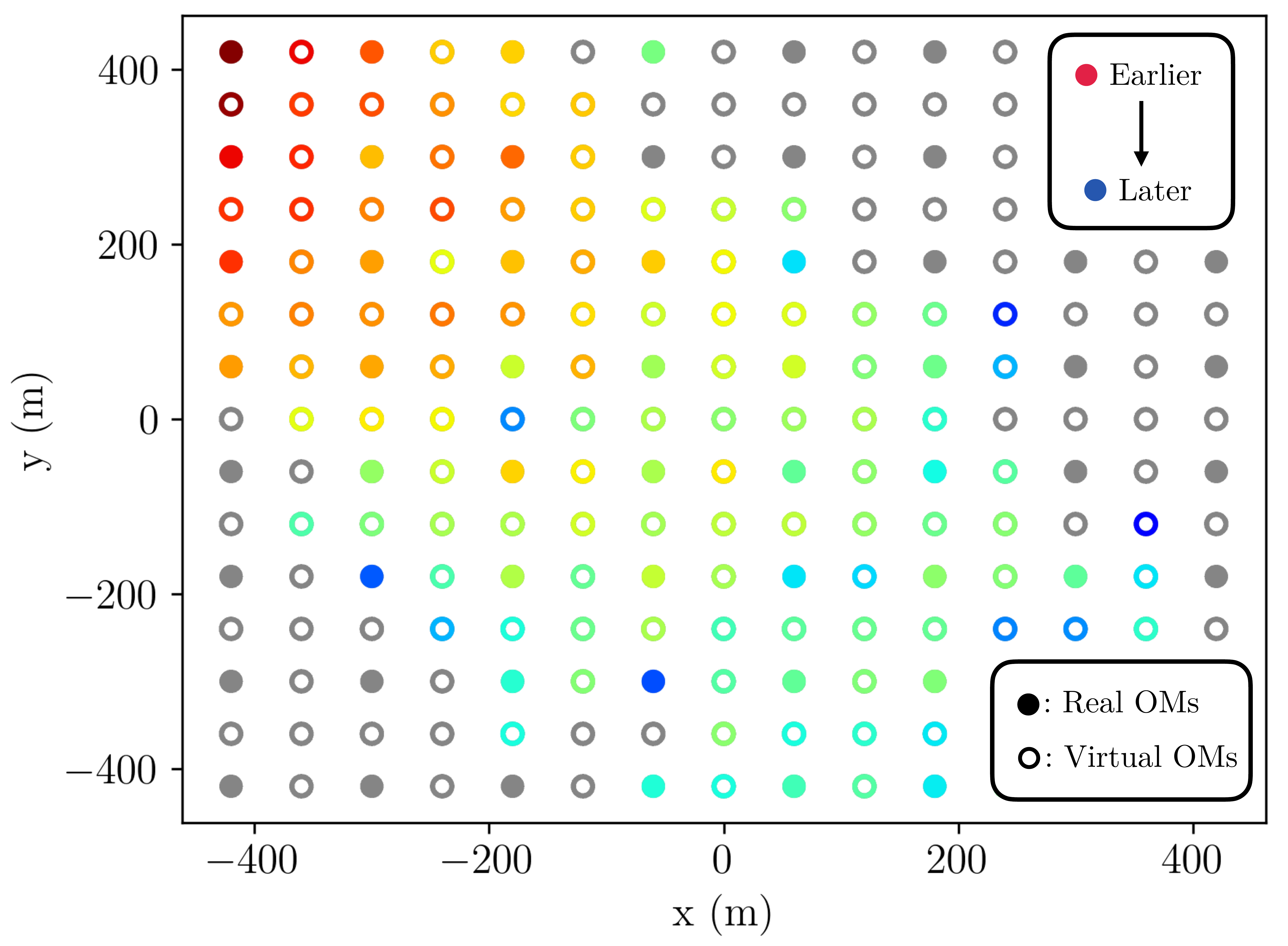}
    \caption{\textbf{\textit{Top-down view of a simulated track event, showing both real and virtual OM strings.}} Color indicates a hit and its timing.}
    \label{fig:masked}
\end{figure}

To produce our dataset, we use the open-source neutrino telescope simulation toolkit \texttt{Prometheus}~\cite{Lazar:2023rol}.
This allows us to define arbitrary detector geometries.
As a benchmark, we use an orthogonal grid of OMs in ice comprising a $15\times15$ array of strings, each with 61~OMs, for a total of 13,705~OMs.
The inter-string spacing is 60\,m, and the OMs vertical spacing is 15\,m.
As input to the network, we mask out every other string, resulting in a $8\times8$ grid of strings with 120\,m spacing.
We note that the virtual OM shadow effect, caused by photons being absorbed when hitting virtual OMs in the denser geometry, is negligible.
Thus, in this particular geometry, characterized by a dense arrangement of 225 strings with 61 OMs per string, we generate $225\times61$ images, as illustrated in \cref{fig:architecture}.

\Cref{fig:masked} shows the top-down view of the virtual and physical string layout.
An example event is shown to demonstrate the effect of masking.
For visualization purposes, only the top OM in each string is shown, and its color indicates the time of the first hit.
Recall that the VAE-based treatment of the hit time distributions described in \cref{sec:arch} allows this method to take advantage of the full timing information of the data, as shown in~\cref{fig:eventdisplay}.

\subsection{Dataset and Training Details}

In this letter, we specifically investigate track-like events, corresponding to muons produced in $\nu_\mu$ charged-current interactions.
The energies of the incident neutrinos are sampled from a power-law distribution with a spectral index of $-1$ between 100\,GeV and 1\,PeV.
We select only events that hit at least eight distinct OMs in the masked geometry.
This procedure results in approximately 500k selected events for the training dataset, and 75k selected events for the test dataset.
We implement all network components using the \texttt{PyTorch} deep-learning framework and use the UNet++ implementation provided by Ref.~\cite{Iakubovskii:2019}.
More details can be found in \cref{app:app_unet}.

\section{\label{sec:results}Results}

The main result of this study is an observed improvement in the angular reconstruction of super-resolved events.
This is presented in \cref{sec:direction_improvement}.
We have also investigated the robustness of our algorithm to uncertainties in the detector medium optical properties, which is discussed in \cref{sec:optical}.

\begin{figure}[t!]
    \centering
    \includegraphics[width=0.48\textwidth]{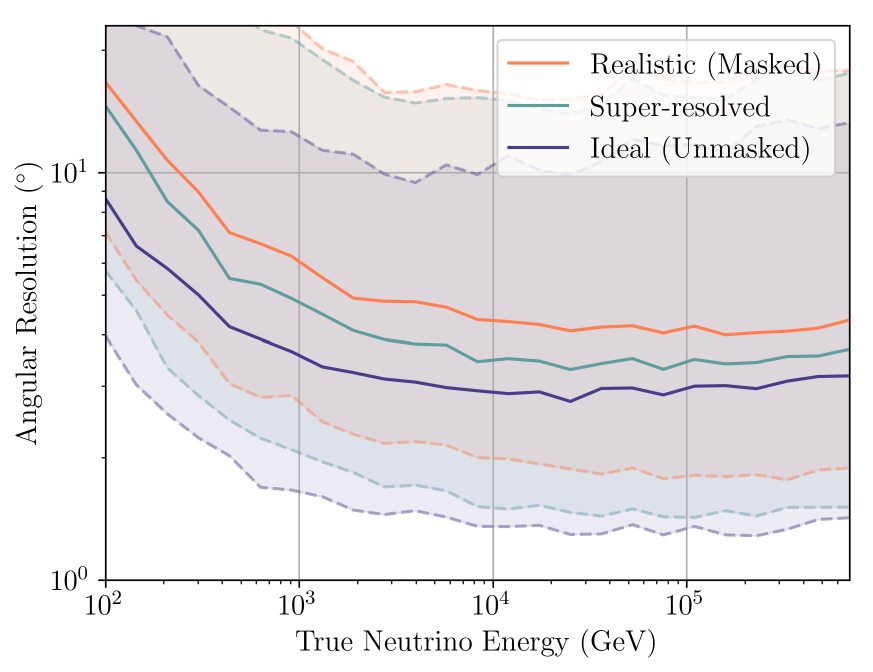}
    \caption{\textbf{\textit{Log-scale angular resolution.}} The median lines are drawn in solid color as a function of the true neutrino energy, produced by a baseline SSCNN method. The 20 and 80 percentiles are denoted by the dashed lines and shaded regions.}
    \label{fig:resolution}
\end{figure}

\subsection{Directional Reconstruction Improvements} \label{sec:direction_improvement}

To assess the performance of the super-resolution network, we run a baseline directional reconstruction method on the masked input events, the super-resolved events produced by the network, and the true unmasked events from the simulation.
We used a sparse submanifold convolutional neural network (SSCNN), as described in Ref.~\cite{PhysRevD.108.063017}; however, we note that the angular resolution improvements should extend to other reconstruction methods.
We train a separate SSCNN model for each of the three datasets.

\Cref{fig:resolution} shows the angular resolution as predicted by the SSCNN algorithm.
As expected, the super-resolved events do not outperform the unmasked events, as the super-resolution network uses the unmasked events as ground truth.
There is noticeable improvement across the whole energy range for the super-resolved dataset, corresponding to about $0.8^\circ$ on average.
The gain is higher in the lower energy bins below 1\,TeV, around $1.3^\circ$ on average. 

The runtime performance is also efficient due to the compression of events into the latent space.
During inference on an NVIDIA A100 GPU, with a batch size of 128, an event takes an average of 0.311 milliseconds on the forward pass.
This is negligible compared to standard angular reconstruction algorithm computational times, which can be tens of milliseconds, and comparable to the SSCNN forward pass time of $\sim 0.1$\,ms~\cite{PhysRevD.108.063017}.

\subsection{Sensitivity to Detector Medium Optical Properties}
\label{sec:optical}

Similar to many ML-based reconstruction methods, the super-resolution process is highly dependent on the simulation and detector properties.
This dependency is particularly significant for super-resolution, as we treat the ``unmasked'' events in the denser geometry from simulations as ground truth.
This naturally poses a challenge to the robustness against varying optical detector properties.
As a case study, we examine the network’s performance on simulation sets with different scattering lengths, $\lambda_s$.
The super-resolution network and subsequent SSCNN reconstruction are both trained on the nominal training sets.
We then generate new test datasets with varied scattering lengths and analyze the reconstruction performance on each set. 

We find that the network is robust to optical medium uncertainties up to approximately 5\%.
This tolerance is sufficient to use this method in current experimental settings.
For example, IceCube reports to have optical property uncertainties at the level of a few percent~\cite{IceCube:2023qua} and are expected to improve with the IceCube Upgrade.
For more details, see \cref{app:sensitivity}.

\section{Conclusions and Future Directions} \label{sec:conclusion}

In this letter, we have introduced a novel technique, super-resolution, for enhancing neutrino telescope event reconstruction.
Our results have demonstrated significant improvements in the angular reconstruction of $\nu_{\mu}$ events in a generic ice-based neutrino telescope.
The approach is also generalizable to different event morphologies and detector types, including water-based neutrino telescopes such as the KM3NeT detector currently being deployed in the Mediterranean Sea. 
Since all neutrino telescope analyses rely on the directional reconstruction of events to pinpoint neutrino sources, the super-resolution technique can enhance the discovery potential of these analyses by improving the performance of existing datasets.

Additionally, the use of a VAE to encode finely-binned timing distributions into a compressed latent space has proven to be an effective strategy.
As a future direction, we aim to explore how to use the VAE to compress neutrino telescope data, enhance and accelerate our reconstructions, and also integrate it into an ML-based event generation model.

As we advance toward constructing larger detectors with larger string spacing, the super-resolution method could play a crucial role in maximizing the scientific output of these experiments.
Furthermore, as we develop more diverse types of detectors, it becomes increasingly important to implement reconstruction methods that work seamlessly across different experiments.

\section*{Acknowledgments}
FY is supported by the National Science Foundation (NSF) through the NSF AI Institute for Artificial Intelligence and Fundamental Interactions.
CAA are supported by the Faculty of Arts and Sciences of Harvard University, the National Science Foundation (NSF), the NSF AI Institute for Artificial Intelligence and Fundamental Interactions, the Research Corporation for Science Advancement, 
NK is supported by the National Science Foundation (NSF) CAREER Award 2239795 and the David and Lucile Packard Foundation.

\bibliography{Upscaling}% Produces the bibliography via BibTeX.

%apsrev4-2.bst 2019-01-14 (MD) hand-edited version of apsrev4-1.bst
%Control: key (0)
%Control: author (8) initials jnrlst
%Control: editor formatted (1) identically to author
%Control: production of article title (0) allowed
%Control: page (0) single
%Control: year (1) truncated
%Control: production of eprint (0) enabled
\begin{thebibliography}{43}%
\makeatletter
\providecommand \@ifxundefined [1]{%
 \@ifx{#1\undefined}
}%
\providecommand \@ifnum [1]{%
 \ifnum #1\expandafter \@firstoftwo
 \else \expandafter \@secondoftwo
 \fi
}%
\providecommand \@ifx [1]{%
 \ifx #1\expandafter \@firstoftwo
 \else \expandafter \@secondoftwo
 \fi
}%
\providecommand \natexlab [1]{#1}%
\providecommand \enquote  [1]{``#1''}%
\providecommand \bibnamefont  [1]{#1}%
\providecommand \bibfnamefont [1]{#1}%
\providecommand \citenamefont [1]{#1}%
\providecommand \href@noop [0]{\@secondoftwo}%
\providecommand \href [0]{\begingroup \@sanitize@url \@href}%
\providecommand \@href[1]{\@@startlink{#1}\@@href}%
\providecommand \@@href[1]{\endgroup#1\@@endlink}%
\providecommand \@sanitize@url [0]{\catcode `\\12\catcode `\$12\catcode `\&12\catcode `\#12\catcode `\^12\catcode `\_12\catcode `\%12\relax}%
\providecommand \@@startlink[1]{}%
\providecommand \@@endlink[0]{}%
\providecommand \url  [0]{\begingroup\@sanitize@url \@url }%
\providecommand \@url [1]{\endgroup\@href {#1}{\urlprefix }}%
\providecommand \urlprefix  [0]{URL }%
\providecommand \Eprint [0]{\href }%
\providecommand \doibase [0]{https://doi.org/}%
\providecommand \selectlanguage [0]{\@gobble}%
\providecommand \bibinfo  [0]{\@secondoftwo}%
\providecommand \bibfield  [0]{\@secondoftwo}%
\providecommand \translation [1]{[#1]}%
\providecommand \BibitemOpen [0]{}%
\providecommand \bibitemStop [0]{}%
\providecommand \bibitemNoStop [0]{.\EOS\space}%
\providecommand \EOS [0]{\spacefactor3000\relax}%
\providecommand \BibitemShut  [1]{\csname bibitem#1\endcsname}%
\let\auto@bib@innerbib\@empty
%</preamble>
\bibitem [{\citenamefont {Arg\"uelles}\ \emph {et~al.}(2024)\citenamefont {Arg\"uelles}, \citenamefont {Halzen},\ and\ \citenamefont {Kurahashi}}]{Arguelles:2024xkx}%
  \BibitemOpen
  \bibfield  {author} {\bibinfo {author} {\bibfnamefont {C.~A.}\ \bibnamefont {Arg\"uelles}}, \bibinfo {author} {\bibfnamefont {F.}~\bibnamefont {Halzen}},\ and\ \bibinfo {author} {\bibfnamefont {N.}~\bibnamefont {Kurahashi}},\ }\href@noop {} {\bibinfo {title} {{From the Dawn of Neutrino Astronomy to A New View of the Extreme Universe}}} (\bibinfo {year} {2024}),\ \Eprint {https://arxiv.org/abs/2405.17623} {arXiv:2405.17623 [hep-ex]} \BibitemShut {NoStop}%
\bibitem [{\citenamefont {Aartsen}\ \emph {et~al.}(2017)\citenamefont {Aartsen} \emph {et~al.}}]{IceCube:2016zyt}%
  \BibitemOpen
  \bibfield  {author} {\bibinfo {author} {\bibfnamefont {M.~G.}\ \bibnamefont {Aartsen}} \emph {et~al.} (\bibinfo {collaboration} {IceCube}),\ }\bibfield  {title} {\bibinfo {title} {{The IceCube Neutrino Observatory: Instrumentation and Online Systems}},\ }\href {https://doi.org/10.1088/1748-0221/12/03/P03012} {\bibfield  {journal} {\bibinfo  {journal} {JINST}\ }\textbf {\bibinfo {volume} {12}}\bibfield  {number} {\bibinfo  {number} { (03)},\ \bibinfo {pages} {P03012}},\ }\bibinfo {note} {[Erratum: JINST 19, E05001 (2024)]},\ \Eprint {https://arxiv.org/abs/1612.05093} {arXiv:1612.05093 [astro-ph.IM]} \BibitemShut {NoStop}%
\bibitem [{\citenamefont {Aartsen}\ \emph {et~al.}(2013{\natexlab{a}})\citenamefont {Aartsen} \emph {et~al.}}]{IceCube:2013cdw}%
  \BibitemOpen
  \bibfield  {author} {\bibinfo {author} {\bibfnamefont {M.~G.}\ \bibnamefont {Aartsen}} \emph {et~al.} (\bibinfo {collaboration} {IceCube}),\ }\bibfield  {title} {\bibinfo {title} {{First observation of PeV-energy neutrinos with IceCube}},\ }\href {https://doi.org/10.1103/PhysRevLett.111.021103} {\bibfield  {journal} {\bibinfo  {journal} {Phys. Rev. Lett.}\ }\textbf {\bibinfo {volume} {111}},\ \bibinfo {pages} {021103} (\bibinfo {year} {2013}{\natexlab{a}})},\ \Eprint {https://arxiv.org/abs/1304.5356} {arXiv:1304.5356 [astro-ph.HE]} \BibitemShut {NoStop}%
\bibitem [{\citenamefont {Aartsen}\ \emph {et~al.}(2013{\natexlab{b}})\citenamefont {Aartsen} \emph {et~al.}}]{IceCube:2013low}%
  \BibitemOpen
  \bibfield  {author} {\bibinfo {author} {\bibfnamefont {M.~G.}\ \bibnamefont {Aartsen}} \emph {et~al.} (\bibinfo {collaboration} {IceCube}),\ }\bibfield  {title} {\bibinfo {title} {{Evidence for High-Energy Extraterrestrial Neutrinos at the IceCube Detector}},\ }\href {https://doi.org/10.1126/science.1242856} {\bibfield  {journal} {\bibinfo  {journal} {Science}\ }\textbf {\bibinfo {volume} {342}},\ \bibinfo {pages} {1242856} (\bibinfo {year} {2013}{\natexlab{b}})},\ \Eprint {https://arxiv.org/abs/1311.5238} {arXiv:1311.5238 [astro-ph.HE]} \BibitemShut {NoStop}%
\bibitem [{\citenamefont {Aartsen}\ \emph {et~al.}(2018)\citenamefont {Aartsen} \emph {et~al.}}]{IceCube:2018cha}%
  \BibitemOpen
  \bibfield  {author} {\bibinfo {author} {\bibfnamefont {M.~G.}\ \bibnamefont {Aartsen}} \emph {et~al.} (\bibinfo {collaboration} {IceCube}),\ }\bibfield  {title} {\bibinfo {title} {{Neutrino emission from the direction of the blazar TXS 0506+056 prior to the IceCube-170922A alert}},\ }\href {https://doi.org/10.1126/science.aat2890} {\bibfield  {journal} {\bibinfo  {journal} {Science}\ }\textbf {\bibinfo {volume} {361}},\ \bibinfo {pages} {147} (\bibinfo {year} {2018})},\ \Eprint {https://arxiv.org/abs/1807.08794} {arXiv:1807.08794 [astro-ph.HE]} \BibitemShut {NoStop}%
\bibitem [{\citenamefont {Abbasi}\ \emph {et~al.}(2022{\natexlab{a}})\citenamefont {Abbasi} \emph {et~al.}}]{IceCube:2022der}%
  \BibitemOpen
  \bibfield  {author} {\bibinfo {author} {\bibfnamefont {R.}~\bibnamefont {Abbasi}} \emph {et~al.} (\bibinfo {collaboration} {IceCube}),\ }\bibfield  {title} {\bibinfo {title} {{Evidence for neutrino emission from the nearby active galaxy NGC 1068}},\ }\href {https://doi.org/10.1126/science.abg3395} {\bibfield  {journal} {\bibinfo  {journal} {Science}\ }\textbf {\bibinfo {volume} {378}},\ \bibinfo {pages} {538} (\bibinfo {year} {2022}{\natexlab{a}})},\ \Eprint {https://arxiv.org/abs/2211.09972} {arXiv:2211.09972 [astro-ph.HE]} \BibitemShut {NoStop}%
\bibitem [{\citenamefont {Abbasi}\ \emph {et~al.}(2023{\natexlab{a}})\citenamefont {Abbasi} \emph {et~al.}}]{IceCube:2023ame}%
  \BibitemOpen
  \bibfield  {author} {\bibinfo {author} {\bibfnamefont {R.}~\bibnamefont {Abbasi}} \emph {et~al.} (\bibinfo {collaboration} {IceCube}),\ }\bibfield  {title} {\bibinfo {title} {{Observation of high-energy neutrinos from the Galactic plane}},\ }\href {https://doi.org/10.1126/science.adc9818} {\bibfield  {journal} {\bibinfo  {journal} {Science}\ }\textbf {\bibinfo {volume} {380}},\ \bibinfo {pages} {adc9818} (\bibinfo {year} {2023}{\natexlab{a}})},\ \Eprint {https://arxiv.org/abs/2307.04427} {arXiv:2307.04427 [astro-ph.HE]} \BibitemShut {NoStop}%
\bibitem [{\citenamefont {Abbasi}\ \emph {et~al.}(2021)\citenamefont {Abbasi} \emph {et~al.}}]{Abbasi:2021ryj}%
  \BibitemOpen
  \bibfield  {author} {\bibinfo {author} {\bibfnamefont {R.}~\bibnamefont {Abbasi}} \emph {et~al.},\ }\bibfield  {title} {\bibinfo {title} {{A Convolutional Neural Network based Cascade Reconstruction for the IceCube Neutrino Observatory}},\ }\href {https://doi.org/10.1088/1748-0221/16/07/P07041} {\bibfield  {journal} {\bibinfo  {journal} {JINST}\ }\textbf {\bibinfo {volume} {16}},\ \bibinfo {pages} {P07041}},\ \Eprint {https://arxiv.org/abs/2101.11589} {arXiv:2101.11589 [hep-ex]} \BibitemShut {NoStop}%
\bibitem [{\citenamefont {Huennefeld}\ \emph {et~al.}(2021)\citenamefont {Huennefeld} \emph {et~al.}}]{IceCube:2021umt}%
  \BibitemOpen
  \bibfield  {author} {\bibinfo {author} {\bibfnamefont {M.}~\bibnamefont {Huennefeld}} \emph {et~al.} (\bibinfo {collaboration} {IceCube}),\ }\bibfield  {title} {\bibinfo {title} {{Combining Maximum-Likelihood with Deep Learning for Event Reconstruction in IceCube}},\ }\href {https://doi.org/10.22323/1.395.1065} {\bibfield  {journal} {\bibinfo  {journal} {PoS}\ }\textbf {\bibinfo {volume} {ICRC2021}},\ \bibinfo {pages} {1065} (\bibinfo {year} {2021})},\ \Eprint {https://arxiv.org/abs/2107.12110} {arXiv:2107.12110 [astro-ph.HE]} \BibitemShut {NoStop}%
\bibitem [{\citenamefont {Abbasi}\ \emph {et~al.}(2022{\natexlab{b}})\citenamefont {Abbasi} \emph {et~al.}}]{IceCube:2022njh}%
  \BibitemOpen
  \bibfield  {author} {\bibinfo {author} {\bibfnamefont {R.}~\bibnamefont {Abbasi}} \emph {et~al.} (\bibinfo {collaboration} {IceCube}),\ }\bibfield  {title} {\bibinfo {title} {{Graph Neural Networks for low-energy event classification \& reconstruction in IceCube}},\ }\href {https://doi.org/10.1088/1748-0221/17/11/P11003} {\bibfield  {journal} {\bibinfo  {journal} {JINST}\ }\textbf {\bibinfo {volume} {17}}\bibfield  {number} {\bibinfo  {number} { (11)},\ \bibinfo {pages} {P11003}},\ }\Eprint {https://arxiv.org/abs/2209.03042} {arXiv:2209.03042 [hep-ex]} \BibitemShut {NoStop}%
\bibitem [{\citenamefont {Kharuk}\ \emph {et~al.}(2023)\citenamefont {Kharuk}, \citenamefont {Safronov}, \citenamefont {Matseiko},\ and\ \citenamefont {Leonov}}]{Kharuk:2023xnl}%
  \BibitemOpen
  \bibfield  {author} {\bibinfo {author} {\bibfnamefont {I.}~\bibnamefont {Kharuk}}, \bibinfo {author} {\bibfnamefont {G.}~\bibnamefont {Safronov}}, \bibinfo {author} {\bibfnamefont {A.}~\bibnamefont {Matseiko}},\ and\ \bibinfo {author} {\bibfnamefont {A.}~\bibnamefont {Leonov}},\ }\bibfield  {title} {\bibinfo {title} {{Machine learning in Baikal-GVD experiment}},\ }\href {https://doi.org/10.22323/1.444.1077} {\bibfield  {journal} {\bibinfo  {journal} {PoS}\ }\textbf {\bibinfo {volume} {ICRC2023}},\ \bibinfo {pages} {1077} (\bibinfo {year} {2023})}\BibitemShut {NoStop}%
\bibitem [{\citenamefont {Eller}(2023)}]{Eller:2023myr}%
  \BibitemOpen
  \bibfield  {author} {\bibinfo {author} {\bibfnamefont {P.}~\bibnamefont {Eller}} (\bibinfo {collaboration} {IceCube}),\ }\bibfield  {title} {\bibinfo {title} {{Public Kaggle Competition ''IceCube -- Neutrinos in Deep Ice''}},\ }in\ \href@noop {} {\emph {\bibinfo {booktitle} {{38th International Cosmic Ray Conference}}}}\ (\bibinfo {year} {2023})\ \Eprint {https://arxiv.org/abs/2307.15289} {arXiv:2307.15289 [astro-ph.HE]} \BibitemShut {NoStop}%
\bibitem [{\citenamefont {De~Sio}(2019)}]{DeSio:2019lcr}%
  \BibitemOpen
  \bibfield  {author} {\bibinfo {author} {\bibfnamefont {C.}~\bibnamefont {De~Sio}} (\bibinfo {collaboration} {KM3NeT}),\ }\bibfield  {title} {\bibinfo {title} {{Machine Learning in KM3NeT}},\ }\href {https://doi.org/10.1051/epjconf/201920705004} {\bibfield  {journal} {\bibinfo  {journal} {EPJ Web Conf.}\ }\textbf {\bibinfo {volume} {207}},\ \bibinfo {pages} {05004} (\bibinfo {year} {2019})}\BibitemShut {NoStop}%
\bibitem [{\citenamefont {Mo}\ \emph {et~al.}(2023)\citenamefont {Mo}, \citenamefont {Hu}, \citenamefont {Li},\ and\ \citenamefont {Xu}}]{Mo:2023rgb}%
  \BibitemOpen
  \bibfield  {author} {\bibinfo {author} {\bibfnamefont {C.}~\bibnamefont {Mo}}, \bibinfo {author} {\bibfnamefont {F.}~\bibnamefont {Hu}}, \bibinfo {author} {\bibfnamefont {L.}~\bibnamefont {Li}},\ and\ \bibinfo {author} {\bibfnamefont {D.}~\bibnamefont {Xu}} (\bibinfo {collaboration} {TRIDENT}),\ }\bibfield  {title} {\bibinfo {title} {{Reconstruction of Track-like Event in TRIDENT Based on a Submanifold Sparse Convolutional Network}},\ }\href {https://doi.org/10.22323/1.444.1205} {\bibfield  {journal} {\bibinfo  {journal} {PoS}\ }\textbf {\bibinfo {volume} {ICRC2023}},\ \bibinfo {pages} {1205} (\bibinfo {year} {2023})}\BibitemShut {NoStop}%
\bibitem [{\citenamefont {Reck}\ \emph {et~al.}(2021)\citenamefont {Reck}, \citenamefont {Guderian}, \citenamefont {Vermari\"en},\ and\ \citenamefont {Domi}}]{Reck:2021zqw}%
  \BibitemOpen
  \bibfield  {author} {\bibinfo {author} {\bibfnamefont {S.}~\bibnamefont {Reck}}, \bibinfo {author} {\bibfnamefont {D.}~\bibnamefont {Guderian}}, \bibinfo {author} {\bibfnamefont {G.}~\bibnamefont {Vermari\"en}},\ and\ \bibinfo {author} {\bibfnamefont {A.}~\bibnamefont {Domi}} (\bibinfo {collaboration} {KM3NeT}),\ }\bibfield  {title} {\bibinfo {title} {{Graph neural networks for reconstruction and classification in KM3NeT}},\ }\href {https://doi.org/10.1088/1748-0221/16/10/C10011} {\bibfield  {journal} {\bibinfo  {journal} {JINST}\ }\textbf {\bibinfo {volume} {16}}\bibfield  {number} {\bibinfo  {number} { (10)},\ \bibinfo {pages} {C10011}},\ }\Eprint {https://arxiv.org/abs/2107.13375} {arXiv:2107.13375 [astro-ph.IM]} \BibitemShut {NoStop}%
\bibitem [{\citenamefont {Choma}\ \emph {et~al.}(2018)\citenamefont {Choma}, \citenamefont {Monti}, \citenamefont {Gerhardt}, \citenamefont {Palczewski}, \citenamefont {Ronaghi}, \citenamefont {Prabhat}, \citenamefont {Bhimji}, \citenamefont {Bronstein}, \citenamefont {Klein},\ and\ \citenamefont {Bruna}}]{IceCube:2018gms}%
  \BibitemOpen
  \bibfield  {author} {\bibinfo {author} {\bibfnamefont {N.}~\bibnamefont {Choma}}, \bibinfo {author} {\bibfnamefont {F.}~\bibnamefont {Monti}}, \bibinfo {author} {\bibfnamefont {L.}~\bibnamefont {Gerhardt}}, \bibinfo {author} {\bibfnamefont {T.}~\bibnamefont {Palczewski}}, \bibinfo {author} {\bibfnamefont {Z.}~\bibnamefont {Ronaghi}}, \bibinfo {author} {\bibnamefont {Prabhat}}, \bibinfo {author} {\bibfnamefont {W.}~\bibnamefont {Bhimji}}, \bibinfo {author} {\bibfnamefont {M.~M.}\ \bibnamefont {Bronstein}}, \bibinfo {author} {\bibfnamefont {S.~R.}\ \bibnamefont {Klein}},\ and\ \bibinfo {author} {\bibfnamefont {J.}~\bibnamefont {Bruna}} (\bibinfo {collaboration} {IceCube}),\ }\href@noop {} {\bibinfo {title} {{Graph Neural Networks for IceCube Signal Classification}}} (\bibinfo {year} {2018}),\ \Eprint {https://arxiv.org/abs/1809.06166} {arXiv:1809.06166 [cs.LG]} \BibitemShut {NoStop}%
\bibitem [{\citenamefont {Yu}\ \emph {et~al.}(2024)\citenamefont {Yu}, \citenamefont {Kamp},\ and\ \citenamefont {Argüelles}}]{Yu-2024}%
  \BibitemOpen
  \bibfield  {author} {\bibinfo {author} {\bibfnamefont {F.~J.}\ \bibnamefont {Yu}}, \bibinfo {author} {\bibfnamefont {N.}~\bibnamefont {Kamp}},\ and\ \bibinfo {author} {\bibfnamefont {C.~A.}\ \bibnamefont {Argüelles}},\ }\href {https://arxiv.org/abs/2410.13148} {\bibinfo {title} {Learning efficient representations of neutrino telescope events}} (\bibinfo {year} {2024}),\ \Eprint {https://arxiv.org/abs/2410.13148} {arXiv:2410.13148 [hep-ex]} \BibitemShut {NoStop}%
\bibitem [{\citenamefont {Jin}\ \emph {et~al.}(2024)\citenamefont {Jin}, \citenamefont {Hu},\ and\ \citenamefont {Arg\"uelles}}]{Jin:2023xts}%
  \BibitemOpen
  \bibfield  {author} {\bibinfo {author} {\bibfnamefont {M.}~\bibnamefont {Jin}}, \bibinfo {author} {\bibfnamefont {Y.}~\bibnamefont {Hu}},\ and\ \bibinfo {author} {\bibfnamefont {C.~A.}\ \bibnamefont {Arg\"uelles}},\ }\bibfield  {title} {\bibinfo {title} {{Two Watts is all you need: enabling in-detector real-time machine learning for neutrino telescopes via edge computing}},\ }\href {https://doi.org/10.1088/1475-7516/2024/06/026} {\bibfield  {journal} {\bibinfo  {journal} {JCAP}\ }\textbf {\bibinfo {volume} {06}},\ \bibinfo {pages} {026}},\ \Eprint {https://arxiv.org/abs/2311.04983} {arXiv:2311.04983 [hep-ex]} \BibitemShut {NoStop}%
\bibitem [{\citenamefont {Zhu}\ \emph {et~al.}(2024)\citenamefont {Zhu}, \citenamefont {Jin},\ and\ \citenamefont {Arg\"uelles}}]{Zhu:2024ubz}%
  \BibitemOpen
  \bibfield  {author} {\bibinfo {author} {\bibfnamefont {T.}~\bibnamefont {Zhu}}, \bibinfo {author} {\bibfnamefont {M.}~\bibnamefont {Jin}},\ and\ \bibinfo {author} {\bibfnamefont {C.~A.}\ \bibnamefont {Arg\"uelles}},\ }\href@noop {} {\bibinfo {title} {{Comparison of Geometrical Layouts for Next-Generation Large-volume Cherenkov Neutrino Telescopes}}} (\bibinfo {year} {2024}),\ \Eprint {https://arxiv.org/abs/2407.19010} {arXiv:2407.19010 [physics.ins-det]} \BibitemShut {NoStop}%
\bibitem [{\citenamefont {Radovic}\ \emph {et~al.}(2018)\citenamefont {Radovic}, \citenamefont {Williams}, \citenamefont {Rousseau}, \citenamefont {Kagan}, \citenamefont {Bonacorsi}, \citenamefont {Himmel}, \citenamefont {Aurisano}, \citenamefont {Terao},\ and\ \citenamefont {Wongjirad}}]{nature-ml}%
  \BibitemOpen
  \bibfield  {author} {\bibinfo {author} {\bibfnamefont {A.}~\bibnamefont {Radovic}}, \bibinfo {author} {\bibfnamefont {M.}~\bibnamefont {Williams}}, \bibinfo {author} {\bibfnamefont {D.}~\bibnamefont {Rousseau}}, \bibinfo {author} {\bibfnamefont {M.}~\bibnamefont {Kagan}}, \bibinfo {author} {\bibfnamefont {D.}~\bibnamefont {Bonacorsi}}, \bibinfo {author} {\bibfnamefont {A.}~\bibnamefont {Himmel}}, \bibinfo {author} {\bibfnamefont {A.}~\bibnamefont {Aurisano}}, \bibinfo {author} {\bibfnamefont {K.}~\bibnamefont {Terao}},\ and\ \bibinfo {author} {\bibfnamefont {T.}~\bibnamefont {Wongjirad}},\ }\bibfield  {title} {\bibinfo {title} {{Machine learning at the energy and intensity frontiers of particle physics}},\ }\href {https://doi.org/10.1038/s41586-018-0361-2} {\bibfield  {journal} {\bibinfo  {journal} {Nature}\ }\textbf {\bibinfo {volume} {560}},\ \bibinfo {pages} {41} (\bibinfo {year} {2018})}\BibitemShut {NoStop}%
\bibitem [{\citenamefont {Li}\ \emph {et~al.}(2023)\citenamefont {Li}, \citenamefont {Fu}, \citenamefont {Winslow}, \citenamefont {Grant}, \citenamefont {Song}, \citenamefont {Ozaki}, \citenamefont {Shimizu},\ and\ \citenamefont {Takeuchi}}]{Li:2022frp}%
  \BibitemOpen
  \bibfield  {author} {\bibinfo {author} {\bibfnamefont {A.}~\bibnamefont {Li}}, \bibinfo {author} {\bibfnamefont {Z.}~\bibnamefont {Fu}}, \bibinfo {author} {\bibfnamefont {L.~A.}\ \bibnamefont {Winslow}}, \bibinfo {author} {\bibfnamefont {C.}~\bibnamefont {Grant}}, \bibinfo {author} {\bibfnamefont {H.}~\bibnamefont {Song}}, \bibinfo {author} {\bibfnamefont {H.}~\bibnamefont {Ozaki}}, \bibinfo {author} {\bibfnamefont {I.}~\bibnamefont {Shimizu}},\ and\ \bibinfo {author} {\bibfnamefont {A.}~\bibnamefont {Takeuchi}},\ }\bibfield  {title} {\bibinfo {title} {{KamNet: An integrated spatiotemporal deep neural network for rare event searches in KamLAND-Zen*}},\ }\href {https://doi.org/10.1103/PhysRevC.107.014323} {\bibfield  {journal} {\bibinfo  {journal} {Phys. Rev. C}\ }\textbf {\bibinfo {volume} {107}},\ \bibinfo {pages} {014323} (\bibinfo {year} {2023})},\ \Eprint {https://arxiv.org/abs/2203.01870} {arXiv:2203.01870 [physics.ins-det]} \BibitemShut {NoStop}%
\bibitem [{\citenamefont {Imani}\ \emph {et~al.}(2024)\citenamefont {Imani}, \citenamefont {Wongjirad},\ and\ \citenamefont {Aeron}}]{Imani:2023blb}%
  \BibitemOpen
  \bibfield  {author} {\bibinfo {author} {\bibfnamefont {Z.}~\bibnamefont {Imani}}, \bibinfo {author} {\bibfnamefont {T.}~\bibnamefont {Wongjirad}},\ and\ \bibinfo {author} {\bibfnamefont {S.}~\bibnamefont {Aeron}},\ }\bibfield  {title} {\bibinfo {title} {{Score-based diffusion models for generating liquid argon time projection chamber images}},\ }\href {https://doi.org/10.1103/PhysRevD.109.072011} {\bibfield  {journal} {\bibinfo  {journal} {Phys. Rev. D}\ }\textbf {\bibinfo {volume} {109}},\ \bibinfo {pages} {072011} (\bibinfo {year} {2024})},\ \Eprint {https://arxiv.org/abs/2307.13687} {arXiv:2307.13687 [hep-ex]} \BibitemShut {NoStop}%
\bibitem [{\citenamefont {Adams}\ \emph {et~al.}(2019)\citenamefont {Adams} \emph {et~al.}}]{MicroBooNE:2018kka}%
  \BibitemOpen
  \bibfield  {author} {\bibinfo {author} {\bibfnamefont {C.}~\bibnamefont {Adams}} \emph {et~al.} (\bibinfo {collaboration} {MicroBooNE}),\ }\bibfield  {title} {\bibinfo {title} {{Deep neural network for pixel-level electromagnetic particle identification in the MicroBooNE liquid argon time projection chamber}},\ }\href {https://doi.org/10.1103/PhysRevD.99.092001} {\bibfield  {journal} {\bibinfo  {journal} {Phys. Rev. D}\ }\textbf {\bibinfo {volume} {99}},\ \bibinfo {pages} {092001} (\bibinfo {year} {2019})},\ \Eprint {https://arxiv.org/abs/1808.07269} {arXiv:1808.07269 [hep-ex]} \BibitemShut {NoStop}%
\bibitem [{\citenamefont {Abratenko}\ \emph {et~al.}(2021)\citenamefont {Abratenko} \emph {et~al.}}]{MicroBooNE:2020yze}%
  \BibitemOpen
  \bibfield  {author} {\bibinfo {author} {\bibfnamefont {P.}~\bibnamefont {Abratenko}} \emph {et~al.} (\bibinfo {collaboration} {MicroBooNE}),\ }\bibfield  {title} {\bibinfo {title} {{Semantic Segmentation with Sparse Convolutional Neural Network for Event Reconstruction in MicroBooNE}},\ }\href {https://doi.org/10.1103/PhysRevD.103.052012} {\bibfield  {journal} {\bibinfo  {journal} {Phys. Rev. D}\ }\textbf {\bibinfo {volume} {103}},\ \bibinfo {pages} {052012} (\bibinfo {year} {2021})},\ \Eprint {https://arxiv.org/abs/2012.08513} {arXiv:2012.08513 [physics.ins-det]} \BibitemShut {NoStop}%
\bibitem [{\citenamefont {Aurisano}\ \emph {et~al.}(2016)\citenamefont {Aurisano}, \citenamefont {Radovic}, \citenamefont {Rocco}, \citenamefont {Himmel}, \citenamefont {Messier}, \citenamefont {Niner}, \citenamefont {Pawloski}, \citenamefont {Psihas}, \citenamefont {Sousa},\ and\ \citenamefont {Vahle}}]{Aurisano:2016jvx}%
  \BibitemOpen
  \bibfield  {author} {\bibinfo {author} {\bibfnamefont {A.}~\bibnamefont {Aurisano}}, \bibinfo {author} {\bibfnamefont {A.}~\bibnamefont {Radovic}}, \bibinfo {author} {\bibfnamefont {D.}~\bibnamefont {Rocco}}, \bibinfo {author} {\bibfnamefont {A.}~\bibnamefont {Himmel}}, \bibinfo {author} {\bibfnamefont {M.~D.}\ \bibnamefont {Messier}}, \bibinfo {author} {\bibfnamefont {E.}~\bibnamefont {Niner}}, \bibinfo {author} {\bibfnamefont {G.}~\bibnamefont {Pawloski}}, \bibinfo {author} {\bibfnamefont {F.}~\bibnamefont {Psihas}}, \bibinfo {author} {\bibfnamefont {A.}~\bibnamefont {Sousa}},\ and\ \bibinfo {author} {\bibfnamefont {P.}~\bibnamefont {Vahle}},\ }\bibfield  {title} {\bibinfo {title} {{A Convolutional Neural Network Neutrino Event Classifier}},\ }\href {https://doi.org/10.1088/1748-0221/11/09/P09001} {\bibfield  {journal} {\bibinfo  {journal} {JINST}\ }\textbf {\bibinfo {volume} {11}}\bibfield  {number} {\bibinfo  {number} { (09)},\ \bibinfo {pages} {P09001}},\ }\Eprint {https://arxiv.org/abs/1604.01444}
  {arXiv:1604.01444 [hep-ex]} \BibitemShut {NoStop}%
\bibitem [{\citenamefont {Domin\'e}\ and\ \citenamefont {Terao}(2020)}]{Domine:2019zhm}%
  \BibitemOpen
  \bibfield  {author} {\bibinfo {author} {\bibfnamefont {L.}~\bibnamefont {Domin\'e}}\ and\ \bibinfo {author} {\bibfnamefont {K.}~\bibnamefont {Terao}} (\bibinfo {collaboration} {DeepLearnPhysics}),\ }\bibfield  {title} {\bibinfo {title} {{Scalable deep convolutional neural networks for sparse, locally dense liquid argon time projection chamber data}},\ }\href {https://doi.org/10.1103/PhysRevD.102.012005} {\bibfield  {journal} {\bibinfo  {journal} {Phys. Rev. D}\ }\textbf {\bibinfo {volume} {102}},\ \bibinfo {pages} {012005} (\bibinfo {year} {2020})},\ \Eprint {https://arxiv.org/abs/1903.05663} {arXiv:1903.05663 [hep-ex]} \BibitemShut {NoStop}%
\bibitem [{\citenamefont {Drielsma}\ \emph {et~al.}(2021)\citenamefont {Drielsma}, \citenamefont {Terao}, \citenamefont {Domin\'e},\ and\ \citenamefont {Koh}}]{Drielsma:2021jdv}%
  \BibitemOpen
  \bibfield  {author} {\bibinfo {author} {\bibfnamefont {F.}~\bibnamefont {Drielsma}}, \bibinfo {author} {\bibfnamefont {K.}~\bibnamefont {Terao}}, \bibinfo {author} {\bibfnamefont {L.}~\bibnamefont {Domin\'e}},\ and\ \bibinfo {author} {\bibfnamefont {D.~H.}\ \bibnamefont {Koh}},\ }\bibfield  {title} {\bibinfo {title} {{Scalable, End-to-End, Deep-Learning-Based Data Reconstruction Chain for Particle Imaging Detectors}},\ }in\ \href@noop {} {\emph {\bibinfo {booktitle} {{34th Conference on Neural Information Processing Systems}}}}\ (\bibinfo {year} {2021})\ \Eprint {https://arxiv.org/abs/2102.01033} {arXiv:2102.01033 [hep-ex]} \BibitemShut {NoStop}%
\bibitem [{\citenamefont {Aiello}\ \emph {et~al.}(2019)\citenamefont {Aiello} \emph {et~al.}}]{KM3NeT:2018wnd}%
  \BibitemOpen
  \bibfield  {author} {\bibinfo {author} {\bibfnamefont {S.}~\bibnamefont {Aiello}} \emph {et~al.} (\bibinfo {collaboration} {KM3NeT}),\ }\bibfield  {title} {\bibinfo {title} {{Sensitivity of the KM3NeT/ARCA neutrino telescope to point-like neutrino sources}},\ }\href {https://doi.org/10.1016/j.astropartphys.2019.04.002} {\bibfield  {journal} {\bibinfo  {journal} {Astropart. Phys.}\ }\textbf {\bibinfo {volume} {111}},\ \bibinfo {pages} {100} (\bibinfo {year} {2019})},\ \Eprint {https://arxiv.org/abs/1810.08499} {arXiv:1810.08499 [astro-ph.HE]} \BibitemShut {NoStop}%
\bibitem [{\citenamefont {Avrorin}\ \emph {et~al.}(2018)\citenamefont {Avrorin} \emph {et~al.}}]{Baikal-GVD:2018isr}%
  \BibitemOpen
  \bibfield  {author} {\bibinfo {author} {\bibfnamefont {A.~D.}\ \bibnamefont {Avrorin}} \emph {et~al.} (\bibinfo {collaboration} {Baikal-GVD}),\ }\bibfield  {title} {\bibinfo {title} {{Baikal-GVD: status and prospects}},\ }\href {https://doi.org/10.1051/epjconf/201819101006} {\bibfield  {journal} {\bibinfo  {journal} {EPJ Web Conf.}\ }\textbf {\bibinfo {volume} {191}},\ \bibinfo {pages} {01006} (\bibinfo {year} {2018})},\ \Eprint {https://arxiv.org/abs/1808.10353} {arXiv:1808.10353 [astro-ph.IM]} \BibitemShut {NoStop}%
\bibitem [{\citenamefont {Agostini}\ \emph {et~al.}(2020)\citenamefont {Agostini} \emph {et~al.}}]{P-ONE:2020ljt}%
  \BibitemOpen
  \bibfield  {author} {\bibinfo {author} {\bibfnamefont {M.}~\bibnamefont {Agostini}} \emph {et~al.} (\bibinfo {collaboration} {P-ONE}),\ }\bibfield  {title} {\bibinfo {title} {{The Pacific Ocean Neutrino Experiment}},\ }\href {https://doi.org/10.1038/s41550-020-1182-4} {\bibfield  {journal} {\bibinfo  {journal} {Nature Astron.}\ }\textbf {\bibinfo {volume} {4}},\ \bibinfo {pages} {913} (\bibinfo {year} {2020})},\ \Eprint {https://arxiv.org/abs/2005.09493} {arXiv:2005.09493 [astro-ph.HE]} \BibitemShut {NoStop}%
\bibitem [{\citenamefont {Aartsen}\ \emph {et~al.}(2021)\citenamefont {Aartsen} \emph {et~al.}}]{IceCube-Gen2:2020qha}%
  \BibitemOpen
  \bibfield  {author} {\bibinfo {author} {\bibfnamefont {M.~G.}\ \bibnamefont {Aartsen}} \emph {et~al.} (\bibinfo {collaboration} {IceCube-Gen2}),\ }\bibfield  {title} {\bibinfo {title} {{IceCube-Gen2: the window to the extreme Universe}},\ }\href {https://doi.org/10.1088/1361-6471/abbd48} {\bibfield  {journal} {\bibinfo  {journal} {J. Phys. G}\ }\textbf {\bibinfo {volume} {48}},\ \bibinfo {pages} {060501} (\bibinfo {year} {2021})},\ \Eprint {https://arxiv.org/abs/2008.04323} {arXiv:2008.04323 [astro-ph.HE]} \BibitemShut {NoStop}%
\bibitem [{\citenamefont {Ye}\ \emph {et~al.}(2022)\citenamefont {Ye} \emph {et~al.}}]{Ye:2022vbk}%
  \BibitemOpen
  \bibfield  {author} {\bibinfo {author} {\bibfnamefont {Z.~P.}\ \bibnamefont {Ye}} \emph {et~al.},\ }\href@noop {} {\bibinfo {title} {{A multi-cubic-kilometre neutrino telescope in the western Pacific Ocean}}} (\bibinfo {year} {2022}),\ \Eprint {https://arxiv.org/abs/2207.04519} {arXiv:2207.04519 [astro-ph.HE]} \BibitemShut {NoStop}%
\bibitem [{\citenamefont {Huang}\ \emph {et~al.}(2023)\citenamefont {Huang}, \citenamefont {Cao}, \citenamefont {Chen}, \citenamefont {Liu}, \citenamefont {Wang}, \citenamefont {You},\ and\ \citenamefont {Qi}}]{Huang:2023mzt}%
  \BibitemOpen
  \bibfield  {author} {\bibinfo {author} {\bibfnamefont {T.-Q.}\ \bibnamefont {Huang}}, \bibinfo {author} {\bibfnamefont {Z.}~\bibnamefont {Cao}}, \bibinfo {author} {\bibfnamefont {M.}~\bibnamefont {Chen}}, \bibinfo {author} {\bibfnamefont {J.}~\bibnamefont {Liu}}, \bibinfo {author} {\bibfnamefont {Z.}~\bibnamefont {Wang}}, \bibinfo {author} {\bibfnamefont {X.}~\bibnamefont {You}},\ and\ \bibinfo {author} {\bibfnamefont {Y.}~\bibnamefont {Qi}},\ }\bibfield  {title} {\bibinfo {title} {{Proposal for the High Energy Neutrino Telescope}},\ }\href {https://doi.org/10.22323/1.444.1080} {\bibfield  {journal} {\bibinfo  {journal} {PoS}\ }\textbf {\bibinfo {volume} {ICRC2023}},\ \bibinfo {pages} {1080} (\bibinfo {year} {2023})}\BibitemShut {NoStop}%
\bibitem [{\citenamefont {Wang}\ \emph {et~al.}(2020)\citenamefont {Wang}, \citenamefont {Chen},\ and\ \citenamefont {Hoi}}]{wang2020deep}%
  \BibitemOpen
  \bibfield  {author} {\bibinfo {author} {\bibfnamefont {Z.}~\bibnamefont {Wang}}, \bibinfo {author} {\bibfnamefont {J.}~\bibnamefont {Chen}},\ and\ \bibinfo {author} {\bibfnamefont {S.~C.}\ \bibnamefont {Hoi}},\ }\bibfield  {title} {\bibinfo {title} {Deep learning for image super-resolution: A survey},\ }\href@noop {} {\bibfield  {journal} {\bibinfo  {journal} {IEEE transactions on pattern analysis and machine intelligence}\ }\textbf {\bibinfo {volume} {43}},\ \bibinfo {pages} {3365} (\bibinfo {year} {2020})}\BibitemShut {NoStop}%
\bibitem [{\citenamefont {Fukuda}\ \emph {et~al.}(2003)\citenamefont {Fukuda} \emph {et~al.}}]{Super-Kamiokande:2002weg}%
  \BibitemOpen
  \bibfield  {author} {\bibinfo {author} {\bibfnamefont {Y.}~\bibnamefont {Fukuda}} \emph {et~al.} (\bibinfo {collaboration} {Super-Kamiokande}),\ }\bibfield  {title} {\bibinfo {title} {{The Super-Kamiokande detector}},\ }\href {https://doi.org/10.1016/S0168-9002(03)00425-X} {\bibfield  {journal} {\bibinfo  {journal} {Nucl. Instrum. Meth. A}\ }\textbf {\bibinfo {volume} {501}},\ \bibinfo {pages} {418} (\bibinfo {year} {2003})}\BibitemShut {NoStop}%
\bibitem [{\citenamefont {Abe}\ \emph {et~al.}(2018)\citenamefont {Abe} \emph {et~al.}}]{Hyper-Kamiokande:2018ofw}%
  \BibitemOpen
  \bibfield  {author} {\bibinfo {author} {\bibfnamefont {K.}~\bibnamefont {Abe}} \emph {et~al.} (\bibinfo {collaboration} {Hyper-Kamiokande}),\ }\href@noop {} {\bibinfo {title} {{Hyper-Kamiokande Design Report}}} (\bibinfo {year} {2018}),\ \Eprint {https://arxiv.org/abs/1805.04163} {arXiv:1805.04163 [physics.ins-det]} \BibitemShut {NoStop}%
\bibitem [{\citenamefont {Abusleme}\ \emph {et~al.}(2022)\citenamefont {Abusleme} \emph {et~al.}}]{JUNO:2021vlw}%
  \BibitemOpen
  \bibfield  {author} {\bibinfo {author} {\bibfnamefont {A.}~\bibnamefont {Abusleme}} \emph {et~al.} (\bibinfo {collaboration} {JUNO}),\ }\bibfield  {title} {\bibinfo {title} {{JUNO physics and detector}},\ }\href {https://doi.org/10.1016/j.ppnp.2021.103927} {\bibfield  {journal} {\bibinfo  {journal} {Prog. Part. Nucl. Phys.}\ }\textbf {\bibinfo {volume} {123}},\ \bibinfo {pages} {103927} (\bibinfo {year} {2022})},\ \Eprint {https://arxiv.org/abs/2104.02565} {arXiv:2104.02565 [hep-ex]} \BibitemShut {NoStop}%
\bibitem [{\citenamefont {Askins}\ \emph {et~al.}(2020)\citenamefont {Askins} \emph {et~al.}}]{Theia:2019non}%
  \BibitemOpen
  \bibfield  {author} {\bibinfo {author} {\bibfnamefont {M.}~\bibnamefont {Askins}} \emph {et~al.} (\bibinfo {collaboration} {Theia}),\ }\bibfield  {title} {\bibinfo {title} {{THEIA: an advanced optical neutrino detector}},\ }\href {https://doi.org/10.1140/epjc/s10052-020-7977-8} {\bibfield  {journal} {\bibinfo  {journal} {Eur. Phys. J. C}\ }\textbf {\bibinfo {volume} {80}},\ \bibinfo {pages} {416} (\bibinfo {year} {2020})},\ \Eprint {https://arxiv.org/abs/1911.03501} {arXiv:1911.03501 [physics.ins-det]} \BibitemShut {NoStop}%
\bibitem [{\citenamefont {Yu}(2024)}]{GithubCode}%
  \BibitemOpen
  \bibfield  {author} {\bibinfo {author} {\bibfnamefont {F.~J.}\ \bibnamefont {Yu}},\ }\href@noop {} {}\bibinfo {howpublished} {\url{https://github.com/felixyu7/nt_mlreco}} (\bibinfo {year} {2024})\BibitemShut {NoStop}%
\bibitem [{\citenamefont {Lazar}\ \emph {et~al.}(2023)\citenamefont {Lazar}, \citenamefont {Meighen-Berger}, \citenamefont {Haack}, \citenamefont {Kim}, \citenamefont {Giner},\ and\ \citenamefont {Arg\"uelles}}]{Lazar:2023rol}%
  \BibitemOpen
  \bibfield  {author} {\bibinfo {author} {\bibfnamefont {J.}~\bibnamefont {Lazar}}, \bibinfo {author} {\bibfnamefont {S.}~\bibnamefont {Meighen-Berger}}, \bibinfo {author} {\bibfnamefont {C.}~\bibnamefont {Haack}}, \bibinfo {author} {\bibfnamefont {D.}~\bibnamefont {Kim}}, \bibinfo {author} {\bibfnamefont {S.}~\bibnamefont {Giner}},\ and\ \bibinfo {author} {\bibfnamefont {C.~A.}\ \bibnamefont {Arg\"uelles}},\ }\bibfield  {title} {\bibinfo {title} {{Prometheus: An Open-Source Neutrino Telescope Simulation}},\ }\href@noop {} {\  (\bibinfo {year} {2023})},\ \Eprint {https://arxiv.org/abs/2304.14526} {arXiv:2304.14526 [hep-ex]} \BibitemShut {NoStop}%
\bibitem [{\citenamefont {Iakubovskii}(2019)}]{Iakubovskii:2019}%
  \BibitemOpen
  \bibfield  {author} {\bibinfo {author} {\bibfnamefont {P.}~\bibnamefont {Iakubovskii}},\ }\href@noop {} {\bibinfo {title} {Segmentation models pytorch}},\ \bibinfo {howpublished} {\url{https://github.com/qubvel/segmentation_models.pytorch}} (\bibinfo {year} {2019})\BibitemShut {NoStop}%
\bibitem [{\citenamefont {Yu}\ \emph {et~al.}(2023)\citenamefont {Yu}, \citenamefont {Lazar},\ and\ \citenamefont {Arg\"uelles}}]{PhysRevD.108.063017}%
  \BibitemOpen
  \bibfield  {author} {\bibinfo {author} {\bibfnamefont {F.~J.}\ \bibnamefont {Yu}}, \bibinfo {author} {\bibfnamefont {J.}~\bibnamefont {Lazar}},\ and\ \bibinfo {author} {\bibfnamefont {C.~A.}\ \bibnamefont {Arg\"uelles}},\ }\bibfield  {title} {\bibinfo {title} {Trigger-level event reconstruction for neutrino telescopes using sparse submanifold convolutional neural networks},\ }\href {https://doi.org/10.1103/PhysRevD.108.063017} {\bibfield  {journal} {\bibinfo  {journal} {Phys. Rev. D}\ }\textbf {\bibinfo {volume} {108}},\ \bibinfo {pages} {063017} (\bibinfo {year} {2023})}\BibitemShut {NoStop}%
\bibitem [{\citenamefont {Abbasi}\ \emph {et~al.}(2023{\natexlab{b}})\citenamefont {Abbasi} \emph {et~al.}}]{IceCube:2023qua}%
  \BibitemOpen
  \bibfield  {author} {\bibinfo {author} {\bibfnamefont {R.}~\bibnamefont {Abbasi}} \emph {et~al.} (\bibinfo {collaboration} {IceCube}),\ }\bibfield  {title} {\bibinfo {title} {{An improved mapping of ice layer undulations for the IceCube Neutrino Observatory}},\ }\href {https://doi.org/10.22323/1.444.0975} {\bibfield  {journal} {\bibinfo  {journal} {PoS}\ }\textbf {\bibinfo {volume} {ICRC2023}},\ \bibinfo {pages} {975} (\bibinfo {year} {2023}{\natexlab{b}})},\ \Eprint {https://arxiv.org/abs/2307.13951} {arXiv:2307.13951 [astro-ph.HE]} \BibitemShut {NoStop}%
\end{thebibliography}%

\appendix
\onecolumngrid

\ifx \standalonesupplemental\undefined
\setcounter{page}{1}

%\newcounter{SIfig}
%\setcounter{SIfig}{1}
\setcounter{figure}{0}

\setcounter{table}{0}
\setcounter{equation}{0}
\fi
\renewcommand{\thepage}{Supplemental Methods and Tables -- S\arabic{page}}

\renewcommand{\figurename}{SUPPL. FIG.}

\renewcommand{\tablename}{SUPPL. TABLE}
\renewcommand{\theequation}{A\arabic{equation}}

\newcounter{SIfig}
\renewcommand{\theSIfig}{SUPPL. FIG. \arabic{SIfig} }

\section{More Details on the UNet} \label{app:app_unet}

This supplement to the main text is intended to provide additional details on the architecture and training methods of the UNet++ used in this letter. 

One of the main strengths of using a VAE to encode timing information on OMs is that the output becomes a fixed-length vector, as defined by the user (the size of the latent dimension). This allows us to effectively compress complex neutrino telescope events into a simple 2D image, where the latent vector is included as part of the features at each pixel. This also enables us to use UNet-type architectures, which has seen much success on 2D images across many scientific disciplines. As mentioned in the main text, we use a latent dimension size of 64.

The features at each OM contains its 3D coordinates, the number of photon hits, and the 64-parameter latent vector containing the timing information. The inputs to UNet++ are masked, while the unmasked events are retained only for loss computation. As is standard for UNet-type architectures, the outputs are also images of the same spatial size. The output has 65 features at each pixel, corresponding to the number of photon hits and the timing latent vector.
MSE loss is used for both the prediction of the number of photon hits and the timing latent vector. The timing latent vector prediction is only trained on OMs that saw light in the event.
We use the \texttt{AdamW} optimizer with an initial learning rate of 0.001, with periodic drops when the validation loss plateaus.
The network is trained for 30 epochs with a batch size of 128 on an NVIDIA A100 GPU.

\section{More Details on Sensitivity to Detector Properties} \label{app:sensitivity}

\begin{figure}[h]
    \centering
    \includegraphics[width=0.5\textwidth]{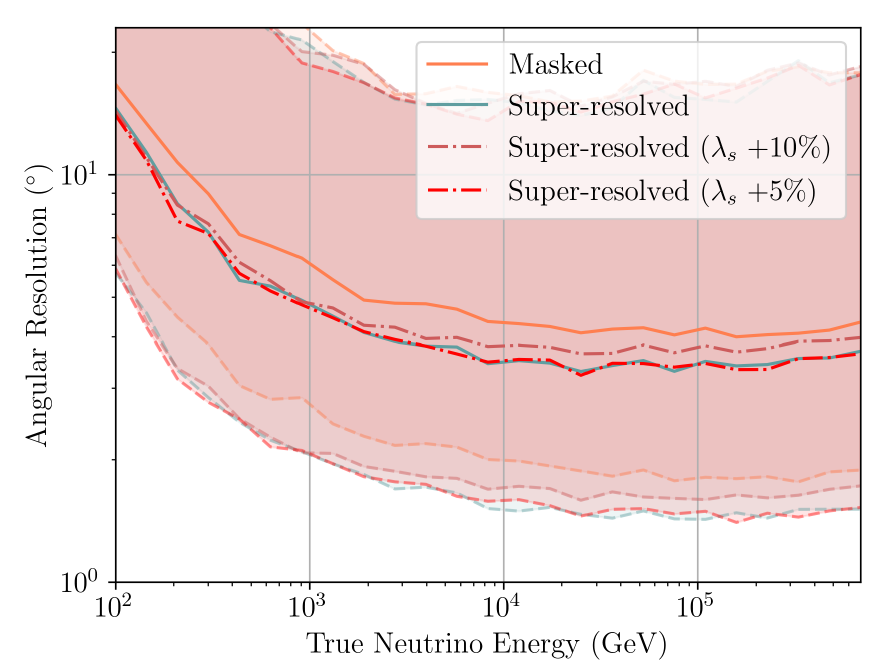}
    \caption{\textbf{\textit{Comparison of angular reconstruction performance when varying scattering length.}} The masked and super-resolved resolutions are obtained from the nominal test dataset, while the dash-dotted median lines are inferred from test datasets with altered scattering lengths.}
    \label{fig:scattering}
\end{figure}

As demonstrated in \cref{fig:scattering}, variations in optical properties at $5\%$ exhibited negligible differences in performance from the nominal test dataset. Increasing the scattering length change to $10\%$ reveals a noticeable decline in reconstruction performance. Interestingly, at lower energies, where reconstruction is more constrained by the lack of light and information, there is minimal impact on performance even with a $10\%$ scattering length change. The performance loss due to the scattering length change is primarily observed at higher energies, where accurate light modeling is more critical.

\end{document}